\documentclass[aps,prd,preprintnumbers,nofootinbib,preprint]{revtex4}
\usepackage{bm}
\usepackage{latexsym}
\usepackage{dcolumn}
\usepackage{amsmath,amsfonts,amssymb}
\usepackage{graphicx,epsfig}
\usepackage{amsthm}
\usepackage{url,hyperref}

\usepackage{slashed}

\newcommand{\be}{\begin{eqnarray}}
\newcommand{\ee}{\end{eqnarray}}
\newcommand{\bea}{\begin{eqnarray}}
\newcommand{\eea}{\end{eqnarray}}

\def\comment#1{}

\usepackage{color}

\definecolor{darkred}{rgb}{.8,0,0}

\definecolor{darkblue}{rgb}{0,0,.7}

\definecolor{darkgreen}{rgb}{0,.7,0}

\begin{document}

%
%
\title{Constraining the deformed dispersion relation with the hydrogen atom 1S-2S transition}
%
%
%
%
%
\author{{Jin Pu}$^{1,2}$}\email[email:~]{pujin@cwnu.edu.cn}
\author{Guo-Ping Li$^{2}$} \email[email:~]{gpliphys@yeah.net}
\author{Qing-Quan Jiang$^{2}$}\email[email:~]{qqjiangphys@yeah.net}
\author{Xiao-Tao Zu$^{1}$\vspace{1ex}}\email[email:~]{xtzu@uestc.edu.cn}
\affiliation{$^1$School of Physics, University of Electronic Science and Technology of China, Chengdu 610054, China\vspace{1ex}}
\affiliation{$^2$College of Physics and Space Science, China West Normal University, Nanchong 637002, China\vspace{1ex}}

%
%
%
%
%
\begin{abstract}
%
%
%
%
%
%
\par\noindent
In this paper, we use the latest results of the ultra-high accuracy 1S-2S transition experiments in hydrogen atom to constrain the forms of the deformed dispersion relation in the nonrelativistic limit. For the leading correction of the nonrelativistic limit, the experiment sets a limit at an order of magnitude for the desired Planck-scale level, thereby providing another example of the Planck-scale sensitivity in the study of the dispersion relation in controlled laboratory experiments. And for the next-to-leading term, bound has two orders of magnitude away from the Planck scale, but it still amounts to the best limit, in contrast to previously obtained bound in the nonrelativistic limit from the cold-atom-recoil experiments.\\
\end{abstract}
\maketitle
%
%
%
%
%
%
\section{Introduction}
\label{intro}
%
%
\par\noindent
Establishing a complete and self-consistent quantum theory of gravity is one of the main challenges in modern physics. Till now, a full understanding of quantum gravity is lacking, but some phenomenological attempts to explore quantum gravity effects have attracted many people's attentions \cite{1,2,3,4,5,6,7,8,9,10,11,12,13,14,15,16,17,18,19,20,21,22,23,24,25,26,27,28,29,30,31,32,33,34,35,36,37,38,39,40,41,42,43,44,45,46,47,48,49,50,51,52,53,54,55,56}. Since most quantum gravity effects are expected to occur at the ultra-high Planck energy scale ($E_p=\sqrt{\hbar c^5/G}\cong 1.2\times 10^{19}GeV$), there are only slight traces on processes that we can approach experimentally. So it is particularly challenging to gain experimental insights into quantum gravity scale. But through tremendous and determined efforts over the last decade, we now have at least a few research lines in quantum gravity phenomenology, in which we have determined that quantum properties of gravity could be studied with the desired Planck-scale sensitivity. For example, due to the ultra-high levels of accuracy of atom interferometry, the cold-atom-recoil experiments have been used to establish meaningful bounds on parameters characterizing quantum gravity effects, and the exceptional sensitivity of the experiments set a limit within a single order of magnitude of the desired Planck-scale level, thereby providing the first example of the Planck-scale sensitivity in the study of the dispersion relation in controlled laboratory experiments \cite{45,57}. In this paper, we attempt to find another example to close to or reach the desired Planck-scale sensitivity by using the latest results of the hydrogen atom 1S-2S transition experiments to constrain the forms of the deformed dispersion relation in the nonrelativistic limit. In \cite{58,59}, quantum gravity corrections to Lamb Shift have computed in the framework of Generalized Uncertainty Principle (GUP) where the accuracy of precision measurement of Lamb Shift of about $1\times 10^{-12}$ leads to the upper bounds on parameters of quantum gravity effects $\beta_0<10^{36}$. In \cite{591}, an upper bound ($\beta_0<10^{34}$) of quantum gravity effects has been obtained by using the high-precision spectrometry of the 1S-2S two photon transition in atomic hydrogen. On the other hand, the progress of frequency conversion technology, such as frequency doubling and frequency division in laser research, makes precision of Lamb Shift experiments in hydrogen atom and deuterium atom ultra high. In Ref.\cite{60,61}, the accuracy of precision measurement of the hydrogen 1S-2S frequency (Lamb Shift experiments) reaches $10^{-15}$. In our case, we use the latest results of the hydrogen atom 1S-2S transition experiments to observe the Planck-scale sensitivity of quantum gravity.
\par
The remainder of this paper is organized as follows. In Sec. \ref{sec2}, we briefly introduce the deformed dispersion relation in nonrelativistic limit. Then, by comparing the results of a detailed calculation of the deformed dispersion relation effects on the 1S-2S transition in hydrogen atom with its accuracy of precision measurement, upper bounds on the parameters of the deformed dispersion relation are obtained in Sec. \ref{sec3}. Sec. \ref{sec4} ends up with some conclusions.
%
%
\section{The deformed dispersion relation in the nonrelativistic limit}
\label{sec2}
%
%
%
\par\noindent

In 2002, Amelino-Camelia has constructed the famous Doubly Special Relativity (DSR), which has two observer-independent constants, i.e. speed of light $c$ and Planck length $L_p$ , of relativity \cite{1}. In the DSR, the deformed dispersion relation naturally leads to the Planck scale departure from Lorentz symmetry, which is referred to as the Lorentz invariance violation of dispersion relations. The related studies were advocating that the general effect of spacetime quantization is the correction of the classical-spacetime dispersion relation between energy $E$ and momentum $p$ of a microscopic particle with mass $m$, usually of the form
\be
E^2=p^2+m^2+p^2(\xi_n\frac{E}{M_p})^n, \label{eeqj1}
\ee
where the speed of light $c$ is set to $1$. These modifications of the dispersion relation over the past decade have been extensively studied by analyzing observational astrophysics data, which of courses involve the ultrarelativistic $(p\gg m)$ system of particle kinematics \cite{33,44,62,63,64,65,66}.
\par
In the nonrelativistic limit $(p\ll m)$, the deformed dispersion relation (\ref{eeqj1}) should be taken the form \cite{45,48}
\be
E\simeq m+\frac{p^2}{2m}+\frac{1}{2M_p}\Big(\xi_1 mp+\xi_2p^2+\xi_3\frac{p^3}{m}\Big).
\label{eq2-1}
\ee
The dispersion relation includes correction terms that are linear in $1/M_p$. In order to really introduce quantum gravity effects in some neighborhood of the Planck scale, the model-dependent dimensionless parameters $\xi_1$, $\xi_2$, $\xi_3$ should have values approximately of order one. And the results from Loop Quantum Gravity \cite{35,68,69,70} and noncommutative geometry \cite{701,702} also have shown that at least some of these parameters should be non-zero. In our case, it is reasonable to use the deformed dispersion relation in the nonrelativistic limit because $p\ll m$( the energy of the electron for $n=1$ state of hydrogen is about $13.6 eV$, but its mass $m\cong0.5\times 10^6eV$).
\par
Unfortunately, just as quantum gravity research usually have challenges, it is also a extremely challenge to translate the theoretically-favoured of values of these parameters of the deformed dispersion relation into a range of possible magnitudes of the effects. From the deformed dispersion relation (\ref{eq2-1}), it find that if the Planck scale is the characteristic scale of quantum gravity effects, the values of these parameters (i.e. $\xi_1$, $\xi_2$, $\xi_3$) should indeed be close to 1, so that the effects of the deformed terms characterized quantum gravity effects is extremely small due to the overall factor $1/M_p$. Although some studies now have shown that the quantum gravity scale may be slightly smaller than the Planck scale, and may even be 3 orders of magnitude smaller than the Planck scale, which is consistent with the the grand unification scale in particle physics \cite{48,71,72}. Therefore these parameters characterizing quantum gravity effects are obtained by 3 orders of magnitude, but the prospect of detectable quantum gravity effects is still very small.
\par
Recently, The Planck-scale sensitivity in the deformed dispersion relation (\ref{eq2-1}) has been studied by using cold atom recoil experiments in Ref. \cite{45}, and the meaningful bounds on the parameters $\xi_1$ and $\xi_2$ have been obtained. The results shown that $\xi_1=-1.8\pm 2.1$ and $|\xi_2|<10^9$, by using the experiments data of Caesium-atom recoil measurements in Ref.\cite{73} and electron-anomaly measurements in Ref.\cite{74}. As discussed above, the range of values of $\xi_1$ indicates that the cold-atom recoil experiments can be considered as the first example of controlled laboratory experiments probing the form of the dispersion relation with sensitivity that is meaningful from a Planck scale perspective. But the bound on parameter $\xi_2$ in the dispersion relation was still a few orders of magnitude away from the Planck scale.
\par
Therefore, our main objective here is to show that the experiment of the ultra-high accuracy 1S-2S transition in hydrogen atom can be used to establish improved bounds on the parameters $\xi_1$ and $\xi_2$ that characterized the nonrelativistic limit of the deformed dispersion relation (\ref{eq2-1}).
%
%
\section{Bounds on the parameters of the deformed dispersion relation}
\label{sec3}
%
%
%
\par\noindent
The hydrogen atom has played a central position in the development of quantum mechanics. As it is the simplest of atoms, it has played an important role for the development and testing of fundamental theories through ever-refined comparisons between experimental data and theoretical predictions, and hydrogen spectroscopy is closely related to the successive advances in the understanding of atomic structure. In recent years, with the advance of experimental technology, the absolute frequency of the 1S-2S transition in atomic hydrogen via two photon spectroscopy has been measured with particularly high precision, so that it can be used to achieve various accurate measurement. For example, the Rydberg constant $R_\infty$ and the proton charge radius have been future improved through the advance of measurement precision of the 1S-2S two photo transition \cite{75}. A value of $R_\infty=10973731.56854(10)m^{-1}$ was obtained. The 1S-2S hydrogen spectroscopy can also be used to search new limits on the drift of fundamental constants \cite{76,77}. Another important application of the 1S-2S two photo transition is used to test electron boost invariance \cite{77}. Inspiring by these achievements with the absolute 1S-2S transition frequency in atomic hydrogen, we think about the possibility of studying quantum gravity effects on the hydrogen atomic spectroscopy.
\par
In our case, we ignore the hyperfine structure, so the hydrogen energy levels are given by \cite{78}
\be
E(n,J,L)=E_{DC}(n,J)+E_{RM}(n,J)+E_{LS}(n,J,L),
\label{eq3-1}
\ee
where $E_{DC}$ and $E_{RM}$ represent the Dirac-Coulomb energy and the energy contributed by the leading recoil corrections due to the finite mass of the nucleus, respectively. These two energy contributions play a major role in the hydrogen energy, which are functions of the Rydberg constant $R_\infty$, the fine structure constant $\alpha$ and the ratio of the electron and nuclear mass $m_e/m_N$. The last term $E_{LS}$ represents the energy contributed by the Lamb shift, which contains the QED corrections and corrections for the finite size and polarizability of the nucleus. Comments on the contributions of hydrogen atoms have been given by in \cite{78,79,80}. In our case, we follow the expression derived by Bethe for the energy level shift. It has been pointed out by Bethe \cite{81} that the displacement of the 2S level of hydrogen observed by Lamb and Retherford \cite{82}, can be simply explained as a shift in the energy of the atom arising from its interaction with the radiation field. Subsequently, by calculating the mean square amplitude of oscillation of an electron coupled to the zero-point fluctuations of the electromagnetic field, the shift of nS energy levels has been given by \cite{83}
\be
\Delta E_n=\frac{4\alpha^2}{3m^2}\big(\ln\frac{1}{\alpha}\big)|\psi_n(0)|^2=\frac{8\alpha^3}{3\pi n^3}\big(\ln\frac{1}{\alpha}\big)\big(\frac{1}{2}\alpha^2m\big)\delta_{l0}.
\label{eq3-2}
\ee
Since the scale of quantum electrodynamic effect is related to the principle quantum number n as $1/n^3$, so the 1S Lamb shift is the largest in atomic hydrogen.
\par
Our main objects here is to expose sensitivity to a meaningful range of values of the parameters $\xi_1$ and $\xi_2$, let us focus on the Planck scale corrections with coefficient $\xi_1$ and $\xi_2$. In the nonrelativistic limit $p\ll m$, since the contribution of the relativistic correction terms to the energy in the relativistic Dirac Hamiltonian is far less than that of the non-relativistic Schrodinger Hamiltonian, we only consider the effect of the Planck scale corrections on the non-relativistic Schrodinger Hamiltonian. Thus, the Planck scale correction terms are regarded as the perturbation terms of the levels energy of hydrogen atom with a well defined quantum Hamiltonian. In the deformed dispersion relation (\ref{eq2-1}), the leading correction and the next-to-leading correction are respectively denoted by Hamiltonian $\hat H'$ and $\hat H''$, where
\be
\hat H'=\xi_1\frac{m}{2M_p}\hat p,~~~~\hat H''=\xi_2\frac{\hat p^2}{2M_p}.
\label{eq3-3}
\ee
Now, we compute the bounds on parameters $\xi_1$ and $\xi_2$ by studying the Planck scale correction of the hydrogen energy levels.
%
%
\subsection{Bounds on the parameter $\xi_1$}
\label{sec3-1}
%
%
\par\noindent
Since the hydrogen atom is spherically symmetric, the Coulomb potential of the hydrogen atom is given by
\be
V(r)=-k/r,
\label{eq3-4}
\ee
where $k=e^2/4\pi\varepsilon_0=\alpha\hbar$, $e$ is electronic charge. To first order, the perturbing Hamiltonian $\hat H'$ shift the energy to
\be
E_n=E_n^{(0)}+\xi_1\frac{m}{2M_p}\langle nlm|\hat p|nlm\rangle,
\label{eq3-5}
\ee
where $E_n^{(0)}=-k/2an^2$, $a$ is the Bohr radius. As discussed above, the 1S Lamb shift is the largest in atomic hydrogen, so we are concerned only with the effects of the Planck scale correction on the shift of 1S energy levels. We have $l=m=0$, and utilize the following to calculate the energy shift: $R_{10}(r)=2a^{-3/2}e^{-r/a}$, $Y_{00}=1/\sqrt {4\pi}$. We derive
\be
<100|\hat p|100>=-i\hbar\langle 100|\frac{\partial}{\partial r}|100\rangle=\frac{i\hbar}{a}.
\label{eq3-6}
\ee
\par
Thus, the shift of energy levels due to the leading correction in the DSR framework is expressed as
\be
\Delta E=|\xi_1\frac{m}{2M_p}\langle 100|\hat p|100\rangle|=\xi_1\frac{m\hbar}{2M_pa}.
\label{eq3-7}
\ee
The additional contribution due to the correction of the parameter $\xi_1$ term in proportion to the original value 1S Lamb shift is given by
\be
\frac{\Delta E}{\Delta E_1}=\xi_1\frac{3\pi m}{8M_p\alpha^4\ln{\frac{1}{\alpha}}}\approx 3.5\times 10^{-15}\xi_1,
\label{eq3-8}
\ee
where some values in Table 1 have been used. As discussed above, if the Planck scale is the characteristic scale of quantum gravity effects, parameter $\xi_1$ should indeed be close to $1$, and then the additional contribution in proportion to the original value (\ref{eq3-8}) is approximately equal to $3.5\times 10^{-15}$. The current accuracy of precision  measurement of the hydrogen 1S-2S transition reach the $4.5\times 10^{-15}$ regime \cite{61}. It interestingly means that the hydrogen 1S-2S transition experiment we here considered can indeed probe the Planck-scale sensitivity on basis of the deformed dispersion relation (\ref{eq2-1}). And, we can finally set out to determine the constraint on the parameter $\xi_1$ by imposing that the corrections are smaller than the experimental error on the value of the hydrogen 1S-2S transition, i.e. $|\xi_1| \leq 1.3$. This estimate is closely related to the degree of coincidence between the physical observation and the theoretical prediction. Since this estimate is determined by using the fine-structure constant $\alpha$ as an input, the uncertainty of this estimate is orders of magnitude above the experiment of the hydrogen 1S-2S transition, i.e. $1.26\times 10^{-7}$.
\par\noindent
\begin{table}
\caption{\bf Quantities used in our calculation}
\label{Tab:Table-1}
\begin{center}
\setlength{\tabcolsep}{1mm}
\linespread{1.5}
\begin{tabular}[t]{|l|c|c|c|}
  \hline
  Quantity  & Value  & Precision  & Source \\
  \hline
$\alpha^{-1}$  & $137.035 999 139(31)$  &  $2.3\times10^{-10}$   &  \cite{84} \\
$m$   & $0.510 998 9461(31) /c^2MeV$   & $6.2\times10^{-9}$  &  \cite{84} \\
$M_p$   & $1.220 910(29)\times10^{19} /c^2GeV$    & $2.3\times10^{-5}$  &  \cite{84} \\
  \hline
\end{tabular}
\end{center}
\end{table}
%
%
\subsection{Bounds on the parameter $\xi_2$}
\label{sec3-2}
%
%
\par\noindent
Following the same steps that we performed above for the correction term with coefficient $\xi_1$, it is easy to verify that the correction term with coefficient $\xi_2$ would produce the following modification of the hydrogen 1S energy levels
\be
\Delta E'=|\langle 100|\hat H''|100\rangle|=\xi_2\frac{1}{2M_p}|\langle 100|\hat p^2|100\rangle|.
\label{eq3-9}
\ee
Using the expression
\be
\hat p^2=2m[\hat H_0+\frac{k}{r}],
\label{eq3-10}
\ee
where $\langle 100|\hat H_0|100\rangle=E_1^{(0)}$, we have
\be
\langle 100|\hat p^2|100\rangle=\frac{mk}{a}=\frac{m\hbar\alpha}{a}.
\label{eq3-11}
\ee
The shift of energy levels due to the next-to-leading correction in the DSR framework is expressed as
\be
\Delta E'=\xi_2\frac{m\alpha\hbar}{2M_pa}.
\label{eq3-12}
\ee
Thus, the additional contribution due to the correction of the parameter $\xi_2$ term in proportion to the original value 1S Lamb shift is given by
\be
\frac{\Delta E'}{\Delta E_1}=\xi_2\frac{3\pi m}{32M_p\alpha^3\ln{\frac{1}{\alpha}}}\approx 2.6\times 10^{-17}\xi_2.
\label{eq3-13}
\ee
\par
According to the current accuracy of precision measurement of the hydrogen 1S-2S transition, the result allow us to establish that $|\xi_2|<10^2$, which means that we indeed can probe the spacetime structure down to length scales of order $10^{-33}m$ ($\sim\xi_2/M_p$). This bound is the best limit on the scenario for the deformation of Lorentz symmetry in the nonrelativistic limit, since previous attempts to constrain the parameter $\xi_2$ is at level $|\xi_2|<10^9$ by using the cold atom recoil experiments \cite{45}. By comparing (\ref{eq3-7}) with (\ref{eq3-12}), it is easy to find that the magnitude of the energy shifts of the hydrogen atom caused by the leading correction term and the next-to-leading correction term differs by the fine structure constant $\alpha$ ($\sim 10^2$). However, in the study of constraining bounds on quantum gravity effects in the deformed dispersion relation by using the cold atom recoil experiment, the leading correction term and the next-to-leading correction term cause the energy correction to differ by a factor $m/(h\nu_\ast+p)$ ($\sim 10^9$)(see details in \cite{45}).
\par
The correction caused by the quadratic term of momentum ($p^2/M_p$) expressed by the parameter $\xi_2$ will indeed become more and more important at high energy. Therefore, some quantum-gravity researchers have used the certain observations in astrophysics to provide Planck-scale sensitivity for some quantum gravity scenarios. These studies have also established meaningful bounds on scenarios with relatively strong ultra-relativistic corrections, such as the proposals of Refs.\cite{85,86,87,88,89} which obtain the bound of the term of order $p^2/M_p$ ($\leq 100$) through gamma-ray bursts (GRBs) and flaring active galactic nuclei (AGNs). And, the bounds of the term of order $p^2/M_p$ ($\leq 1$) could be obtained by using neutrino events detected by IceCube Collaboration in Refs.\cite{90,91,92,93,94}. It means that our bound is two orders of magnitude higher than these meaningful bounds established in astrophysics observations. Thus, the hydrogen 1S-2S transition experiments can be considered to be able to investigate the desired Planck scale sensitivity.
%
%
\section{Conclusion}
\label{sec4}
%
%
%
\par\noindent
We use the latest results of the ultra-high accuracy 1S-2S transition experiments in hydrogen atom to establish upper bounds on parameters $\xi_1$ and $\xi_2$ characterizing the nonrelativistic limits of the deformed dispersion relation. The results show that the exceptional sensitivity of the experiments sets a limit on parameter $\xi_1$ within a single order of magnitude of the desired Planck-scale level, thereby providing another example of the Planck-scale sensitivity in the study of the dispersion relation in controlled laboratory experiments. At the same time, bound of parameter $\xi_2$ has two orders of magnitude away from the Planck scale, but it still amounts to the best limit, in contrast to previously obtained bounds in the nonrelativistic limit from the cold-atom-recoil experiments \cite{45,57}. We can expect that, as the hydrogen atom 1S-2S transition experiments continue to improve, more stringent bounds on parameters $\xi_1$ and $\xi_2$ could be found in the near future.
%
%
\section{Acknowledgements}
\par\noindent
This work is supported by the Program for NCET-12-1060, by the Sichuan Youth Science and Technology Foundation with Grant No. 2011JQ0019, and by FANEDD with Grant No. 201319, and by the Innovative Research Team in College of Sichuan Province with Grant No. 13TD0003, and by Ten Thousand Talent Program of Sichuan Province, and by Sichuan Natural Science Foundation with Grant No. 16ZB0178, and by the starting funds of China West Normal University with Grant No.17YC513 and No.17C050.
%
%

%
%
\end{document}